\documentstyle[11pt,emulateapj,epsfig,psfig]{article} % <=== ApJformat
\newcommand{\be}{\begin{equation} }
\newcommand{\ee}{\end{equation} }
\newcommand{\bea}{\begin{eqnarray} }
\newcommand{\eea}{\end{eqnarray} }
\newcommand{\kes}{\kappa_{es}}

\lefthead{Zane et al.}
\righthead{PROTON CYCLOTRON FEATURES}

\begin{document}

\title{Proton Cyclotron Features in Thermal Spectra of Ultra-magnetized
Neutron Stars}

\author{S. Zane\altaffilmark{1}, R. Turolla\altaffilmark{2},
L. Stella\altaffilmark{3}, A. Treves\altaffilmark{4}}

\altaffiltext{1}{Mullard Space Science Laboratory, University College
London,  Holmbury St. Mary, Dorking, Surrey, RH5 6NT, UK;
e-mail: sz@mssl.ucl.ac.uk}
\altaffiltext{2}{Dipartimento di Fisica, Universit\`a di Padova,
Via Marzolo 8, I-35131 Padova, Italy;
e-mail: turolla@pd.infn.it}
\altaffiltext{3}{Osservatorio Astronomico di Roma,
via dell'Osservatorio 2, I-00040 Monte Porzio Catone, Roma, Italy;
e-mail: stella@coma.mporzio.astro.it }
\altaffiltext{4}{Dipartimento di Scienze, Universit\`a dell'Insubria, Via
Valleggio 11, Como, Italy; e-mail: treves@mib.infn.it}

\begin{abstract}

A great deal of interest has been recently raised in connection
with the possibility that soft $\gamma$-ray repeaters (SGRs) and
anomalous X-ray pulsars (AXPs) contain {\em magnetars},
young neutron stars endowed with magnetic fields $\gtrsim 10^{14}$ G.
In this paper we calculate thermal spectra from
ultra-magnetized neutron stars for values of the
luminosity and magnetic field believed to be relevant to
SGRs and AXPs. Emergent spectra
are found to be very close to a blackbody at the star effective
temperature and exhibit a distinctive absorption feature at the proton
cyclotron energy $E_{c,p}\simeq 0.63 (B/10^{14}\, {\rm G})$ keV. The
proton cyclotron features (PCFs) are conspicuous (equivalent width of up to
many hundreds eV) and relatively broad ($\Delta E/E \sim 0.05-0.2$).
The detection of the PCFs is well within the capabilities of present X-ray
spectrometers, like the HETGS and METGS on board Chandra. Their observation
might provide decisive evidence in favor of the existence of magnetars.

\end{abstract}


\section{Introduction}\label{sec-intro}

Over the last few years increasing observational evidence
has gathered in favor of the existence of ultra-magnetized neutron stars
(NSs) with surface field $B\gtrsim 10^{14}$ G. The existence of these {\em
magnetars} was first proposed nearly
10 years ago by Thompson \& Duncan (\cite{td93:1993}) who showed that,
during the core collapse, convective motions can strongly amplify
the seed magnetic field.

Magnetars are unique among compact stars, in that their
luminosity should be provided by the dissipation of magnetic
energy. Like conventional radio-pulsars, however, they will also
lose rotational energy according to the magnetic dipole formula.
Owing to their high magnetic field, magnetars are expected to
spin down at a high rate, $\dot P \approx 10^{-11}(B/10^{14}\,
{\rm G})^2/P$ ss$^{-1}$. It was the detection of pulsations with
a secular spin-down in this range in two soft $\gamma$-repeaters
(SGRs; Kouveliotou et al. \cite{ketal98:1998};
\cite{ketal99:1999}) that for the first time suggested the
association of these sources with ultra-magnetized NSs. Besides
their bursting activity, SGRs show also persistent X-ray emission
with luminosities $L\approx 10^{34}-10^{36} \ {\rm erg\,s^{-1}}$.
X-ray spectra are usually power-law like, but in one case (SGR
1900+14) clear evidence was found for a thermal spectral
component at $kT\sim 0.5$ keV (Woods et al. \cite{woo99:1999}).
This is believed to originate from the NS surface radiating away
the dissipated magnetic energy (see e.g. Thompson \cite{th00:2000}
for a recent review).

Magnetars have been also invoked to explain the characteristics
of another enigmatic class of Galactic sources, the so-called
anomalous X-ray pulsars (AXPs). These pulsars share properties
that are much at variance with those of canonical X-ray pulsators
(Mereghetti \& Stella \cite{meste95:1995}).
Among these are: i) a narrow interval of pulse periods (6--12 s);
ii) very soft X-ray spectra, often with a blackbody-like
component at $kT\sim 0.4$--0.6 keV; iii) relatively low X-ray
luminosities ($\sim 10^{35} \ {\rm erg\,s^{-1}}$); iv) spin-down
trends at relatively large and stable rates ($\sim 0.6-30\times 10^{-12} \
{\rm s\,s}^{-1}$); v) absence of massive companion
stars and vi) association with supernova remnants (see e.g. Mereghetti 2000
for a review).

Though it is now widely recognized that AXPs are rotating NSs,
two different mechanisms for powering their X-ray emission have
been discussed: accretion (from a very low mass companion, fall
back or a fossil disk, Van Paradijs, Taam, \& Van den Heuvel
\cite{vp:1995}; Chatterjee, Hernquist, \& Narayan \cite{ch:2000})
and magnetic energy dissipation in ultra-magnetized, isolated
NSs. If AXPs are young, hot NSs giving off thermal radiation,
magnetic field decay should provide crustal heating for a long
enough time. Possible decay scenarios have been analyzed by Colpi,
Geppert \& Page (\cite{cpd00:2000}) and led to the conclusion
that AXPs are $\approx 10^4$ yr old with a present $B$-field of
$\approx 10^{14}$ G. The latter figure is in the range of that
inferred from SGRs spin-down rates. The many similarities between
AXPs and SGRs strengthen the idea that the two classes of sources
are powered by the same mechanism, i.e. magnetic field dissipation
(e.g. Thompson \cite{th00:2000}).

The transfer of radiation in strongly magnetized atmospheres has been
widely investigated in connection with X-ray emission from cooling
and accreting neutron stars (see e.g. Shibanov et al.
\cite{sh92:1992}; Rajagopal, \& Romani \cite{rr96:1996};
Pavlov et al. \cite{pztn96:1996}; Zane, Turolla \& Treves \cite{ztt00:2000},
herafter ZTT).
Up to now, however, little attention was given to emission from
ultra-magnetized neutron stars with surface fields $\gtrsim 10^{13}$ G.
Detailed transfer calculations
to assess the emergent spectrum from atmospheres around ultra-magnetized
NSs are therefore needed.
Model fits to the X-ray spectra of AXPs and SGRs indicate the presence of
a power-law, high-energy component. In
addition to this, thermal emission has been
detected in all AXPs but one, and in SGR 1900+14 (see Mereghetti
\cite{me00:2000} and references therein).

In this paper we present numerical models of thermal emission
from NSs for a range of luminosity, $L\sim 10^{34}-10^{36} \ {\rm
erg\,s}^{-1}$, and magnetic field, $B\sim 10^{14}-10^{15}$ G,
which are thought typical of SGRs and AXPs. Emergent spectra,
computed solving the full (angle and energy dependent) transfer
equations, are nearly Planckian, have negligible hardening, and
exhibit a strong absorption feature at the proton cyclotron
resonance, $E_{c,p}\simeq 0.63 y_G(B/10^{14}\, {\rm G})$ keV,
where $y_G= \sqrt{1-2GM/c^2R}$ is the gravitational redshift
factor ($M$ and $R$ are the star mass and radius, respectively).
The predicted equivalent width of the proton cyclotron features
(PCFs) is in the range $0.03$--1 keV and the relative width
$FWZI/E_{c,p}\approx 0.7$ (here $FWZI$ is the full width at zero
intensity). The detection of PCFs is definitely within the range
of Chandra spectrometers. Their actual observation in the X-ray
spectra of SGRs and/or AXPs would provide a direct measurement of
the surface magnetic field and a remarkable confirmation of the
magnetar model.

\section{Model Atmospheres}\label{sec-model}

In this paper  we follow closely the approach presented in ZTT,
to which we refer for details. In
particular we consider  a magnetized, non-degenerate, pure
hydrogen plasma in thermal equilibrium at temperature $T$ in which
the dominant radiative processes are free-free
emission/absorption and Thomson scattering; non-conservative
scattering has negligible effect on the transfer of radiation but becomes
important in the energy balance of the outer layers
(see below). We approximate the atmosphere as a plane-parallel
slab with the magnetic field ${\bf B}$ parallel to the normal
${\bf n}$ (the $z$-axis). This is reasonable for magnetized
accretors, but it becomes questionable in treating emission from the
entire star surface, as in the case of cooling neutron stars. The
effects induced on the emerging spectrum by including surface
elements away from the magnetic poles are therefore discussed in section
\ref{sec-results}. Although we are mainly interested in
thermal emission from magnetars, ultra-magnetized NSs
might also accrete the interstellar medium, as recently suggested by
Rutledge (\cite{rut01:2001}). In the light
of this, and for the sake of completeness, we consider here both the case of
accreting and cooling atmospheres.

For photon frequencies in the range $\omega_p<\omega<
\omega_{c,e}$, where $\omega_p$ and $\omega_{c,e} = eB/m_ec$ are
the plasma and the electron cyclotron frequency respectively, the
semi-transverse approximation can be used and the transfer of
radiation through the magneto-active medium is described in terms
of the two normal modes (ZTT and references therein). Atmospheric models are
constructed by solving
the transfer equations for the normal modes coupled to the
hydrostatic equilibrium and the energy equation. Since the
luminosity $L$ is well below the Eddington limit, the hydrostatic
balance gives immediately the density as a function of the
scattering depth $\tau=\int_\infty^z \kes\rho\, dz$

\be\label{dens} \rho = \frac{GMm_p}{2
y_G^2R^2\kes}\frac{\tau}{kT(\tau)} \ee (here $\kes=0.4 \ {\rm cm^2\,g^{-1}}$ is the
electron scattering opacity for an unmagnetized medium; see ZTT).
The energy balance simply states that the net radiative cooling
must equate the energy $W_H$ released in the atmosphere by
non-radiative processes, like the bombardment of impinging ions;
for cooling atmospheres $W_H=0$. Including Compton
heating/cooling the energy equation becomes

\be \label{en}
k_P \frac{aT^4}{2} - \sum_{i=1}^2 k_{am}^{(i)}U^{(i)} + \left (\Gamma -
\Lambda \right)_C
=\frac{W_H}{c\kes}
\ee
where $U^{(i)}$ is the radiation energy density of mode $i$, $k_P$,
$k^{(i)}_{am}$ are defined in strict analogy with the Planck and absorption
mean opacities in the unmagnetized case and
$(\Gamma - \Lambda)_C$ is the Compton heating/cooling rate in a magnetized
plasma (see again ZTT).

\newpage
\section{Results}\label{sec-results}

\begin{figure}
\epsfxsize=17.truecm
\centerline{{\epsfbox{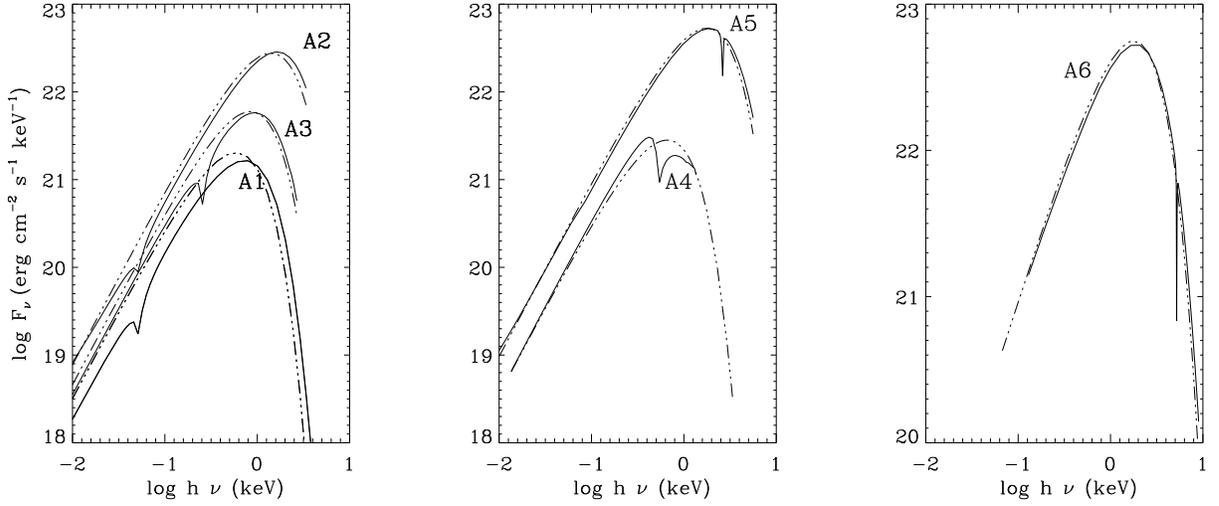}}}
\caption{Emergent spectra (solid lines) for the models presented in table
1, together with the blackbody at the NS  effective temperature
(dash-dotted lines). Energies have been red-shifted at Earth while
the monochromatic flux (on the vertical axis) is evaluated at the star surface.
\label{fig_spec}}
\end{figure}

Following the method outlined in the previous section, several
model atmospheres have been computed for different values of the
magnetic field strength and luminosity. As expected (see
ZTT; Treves et al.
\cite{treves00:2000}), cooling and accreting atmospheres turn out
to produce essentially the same spectrum at X-ray energies, so we
do not make further distinction between them. It should be kept
in mind, however, that a cooling model characterized by a
luminosity $L$ is equivalent to an accreting one with luminosity
$f_AL<L$, where $f_A$ is the fraction of the emitting area. All
models have been calculated for $M= 1M_\odot$, $R=9$ km and are
labeled by the value of the luminosity measured by an observer at
infinity, assuming emission from the entire surface (i.e.
$f_A=1$). We have explored the ranges $10^{13}\, {\rm G}\leq B\leq
10^{15}\, {\rm G}$ and $10^{34}\, {\rm erg\,s}^{-1}\lesssim
L\lesssim 10^{36}\, {\rm erg\,s}^{-1}$ which are believed typical of
magnetars. The main parameters of the computed
models are summarized in table \ref{tab1} and emerging spectra are shown
in figure \ref{fig_spec}. In
all cases, with the exception of model A4, the spectra are nearly
Planckian in shape and show a small hardening with respect to the
blackbody at the star effective temperature $T_{eff}$. As it was found in
previous investigations (e.g. Shibanov et al. \cite{sh92:1992};
ZTT), the spectral
hardening decreases with increasing field strength and almost
vanishes for $B \gtrsim  10^{14}$ G. The most
prominent spectral signature is the absorption feature at the
proton cyclotron resonance $E_{c,p}\simeq 0.63 y_G (B/10^{14}\, {\rm G})$
keV which falls in the soft-medium X-rays for $B \sim
10^{14}$--$10^{15}$ G; the electron cyclotron line
is at $\simeq 1.2 (B/10^{14}\, {\rm G})$ MeV. The line equivalent
width, $EW$, and the full width at zero intensity relative to the line
center, $\Delta E/E \equiv FWZI/E$, are reported
in table \ref{tab1}. Two main effects contribute to the line profile:
the intrinsic resonance in the
magnetic absorption coefficients (this essentially produces Fraunh\"ofer
lines) and mode crossing at the mode
collapse points. The latter is amplified when collapse points
induced by vacuum effects fall near the resonance and in the photospheric
region (see also Shibanov et al. \cite{sh92:1992}). The large
modification in the continuum in model A4, for instance, is partially due
to mode-crossing which, for these values of the parameters, occurs close
to the Wien peak.

\begin{deluxetable}{cccccc}
\tablecolumns{6}
\tablewidth{0pt}
\tablecaption{Model Parameters\label{tab1}}
\tablehead{
\colhead{Model}&
\colhead{$B$} &
\colhead{$L$} &
\colhead{$E_{c,p}^a$} &
\colhead{$EW$} &
\colhead{$\Delta E/E$} \\
\colhead{} &
\colhead{$10^{13}$ G} &
\colhead{$10^{34}$ erg s$^{-1}$} &
\colhead{keV} &
\colhead{keV} &
\colhead{}
}
\startdata
 A1 & 1 & 2 & 0.06 & 0.01 & 0.67 \\
 A2 & 1 & 70 & 0.06 & 0.01 & 0.47 \\
 A3 & 5 & 9 & 0.32 &0.05& 0.31 \\
 A4 & 10 & 3 & 0.63 & -- & -- \\
 A5 & 50 & 170 & 3.15 & 0.10 & 0.10 \\
 A6 & 100 & 180 & 6.3 & 0.11 & 0.06 \\
\tablecomments{The line energy is not corrected for the gravitational
red-shift; the corresponding value at Earth is lower by a factor
$y_G \sim 0.8$.}
\enddata
\end{deluxetable}

As discussed earlier, the assumption of a constant magnetic field
breaks down when emission comes from the entire star surface.
Even for a simple dipolar field, ${\bf B}$ changes in both magnitude
and direction along the surface and this will
produce a broadening of the PCF. Moreover, large
$B$-fields introduce a meridional temperature variation that
makes the emergent spectrum dependent of the viewing angle. A
thorough modeling would require a genuine two-dimensional
transfer calculation, which is still beyond the capabilities of
present numerical codes. However, an estimate of the expected
broadening can be obtained with a simple, approximated
computation. To do this, we assume a relativistic magnetic dipole
in the Schwarzschild space-time, aligned with the rotational axis
(see e.g. Pavlov \& Zavlin \cite{pz00:2000})
\begin{equation}
B = {B_p \over 2} \left [ \left ( 4 - f^2 \right ) \cos^2 \chi +
f^2 \right ]^{1/2}
\end{equation}
where $B_p$ is the field strength at the magnetic pole ($\chi =
0$) and $f\simeq 1.1$ accounts for gravitational effects.
The surface temperature
profile is taken from Possenti, Mereghetti \& Colpi (\cite{pmc96:1996})
\begin{equation} \label{tsu}
T = T_{eff}
\left \{ \frac{ K
+ \left ( 4 - K \right ) \cos^2 \chi}{\left [ 1 - 0.47 \left ( 1 - K
\right )\right ]\left(1 + 3 \cos^2 \chi\right) } \right\}^{1/4}
\end{equation}
where $K$ is the ratio of the coefficients of thermal
conductivity orthogonal and parallel to the field. The exact value of
$K$ is not very important: for $B \gtrsim 10^{13}$
G it is $K  \lesssim 10^{-4}$, and beyond this value the temperature
profile depends weakly on $K$.
The flux emitted by the surface
element at co-latitude $\chi$ is evaluated assuming thermal diffusion for
both modes
\begin{equation}
d F_\nu = 2\pi\sum_{i=1}^2 \frac{
D^{(i)}}{2} \frac{d B_\nu[\,T(\chi)]}{d \tau }\sin\chi\,d\chi
\label{fdiff}
\end{equation}
where $ B_\nu$ is the blackbody function. The diffusion
coefficients $D^{(i)}$ are obtained by angle-averaging the reciprocal
of the total opacity over all photon directions (see Shibanov et
al. \cite{sh92:1992}). They depend on
$\mu_B = {\bf n}\cdot{\bf B}/B$; here ${\bf n} = {\bf r}/r$ is the ``local''
normal to the surface.
The total flux is then computed by integrating equation
(\ref{fdiff}) over the entire star surface, with the aid of
expression
\begin{equation}
\mu_B^2 = \frac{1}{ 4 - f^2} \left [ 4 - \frac{4 f^2}{\left ( 4
- f^2 \right ) \cos^2 \chi + f^2 } \right ]
\end{equation}
which relates $\mu_B$ to $\chi$ (see again Pavlov \& Zavlin \cite{pz00:2000}).

\begin{figure}

\centerline{
\epsfxsize=6.truecm
\centerline{{\epsfbox{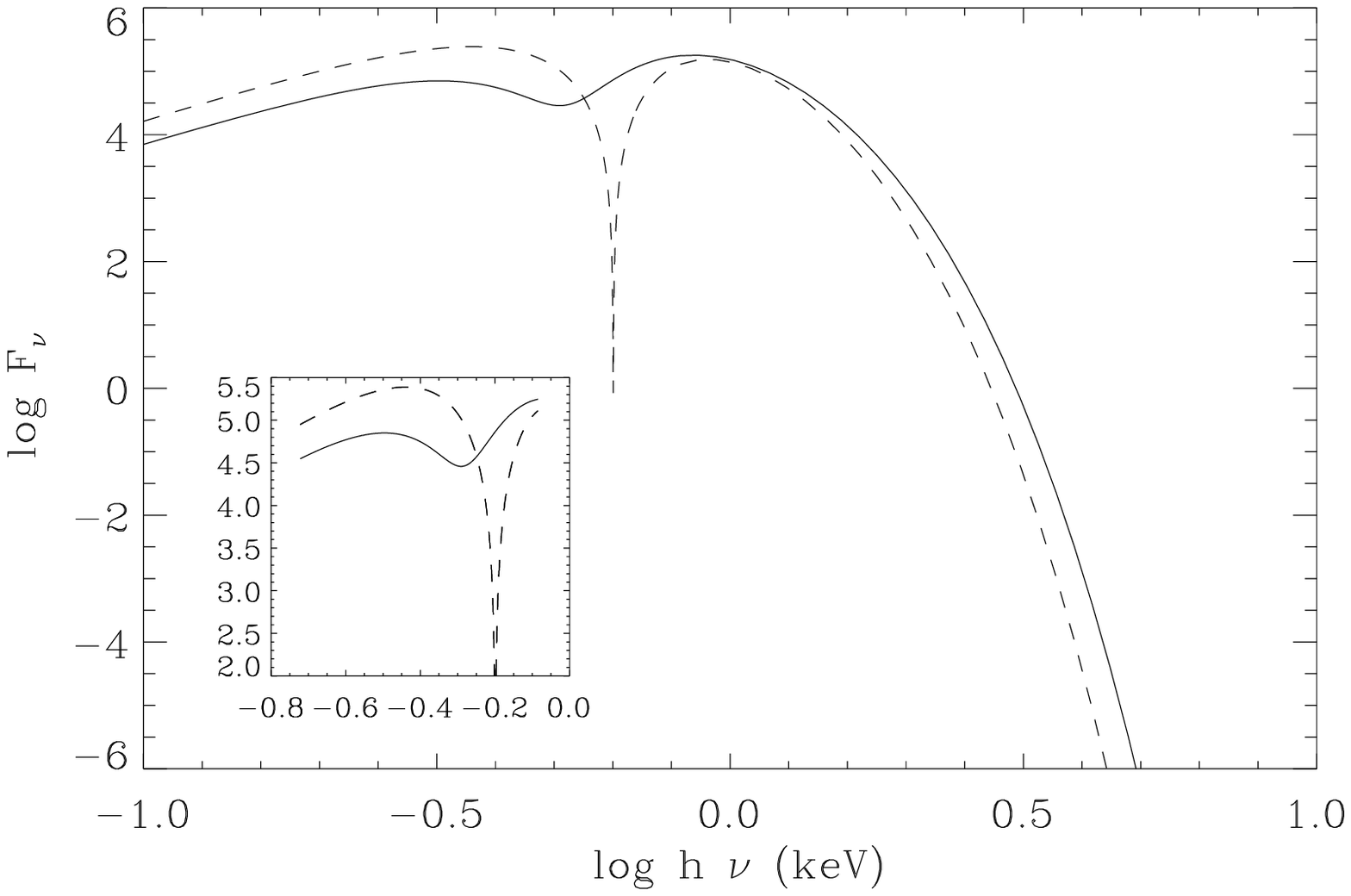}}
{\epsfbox{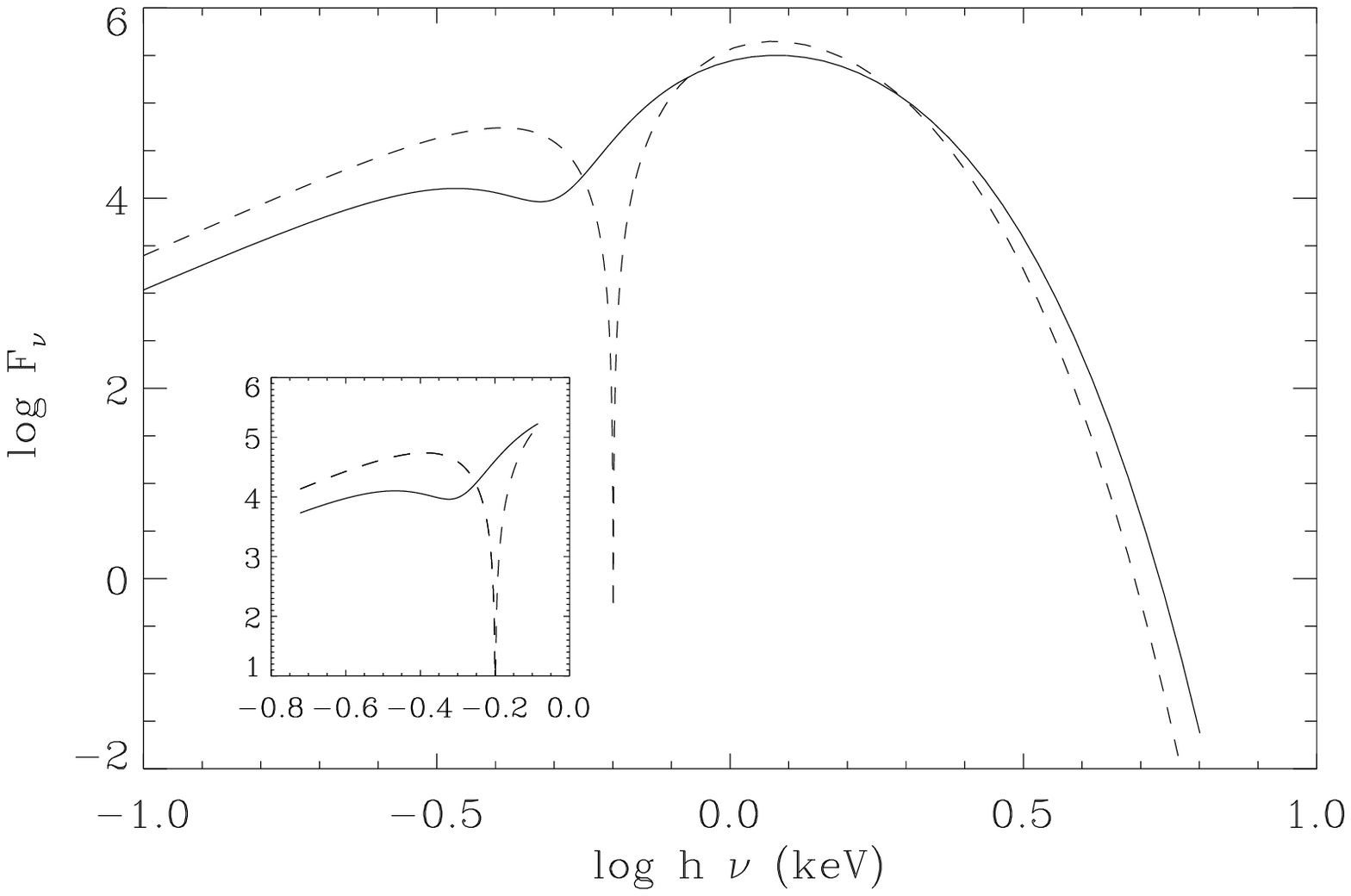}}
{\epsfbox{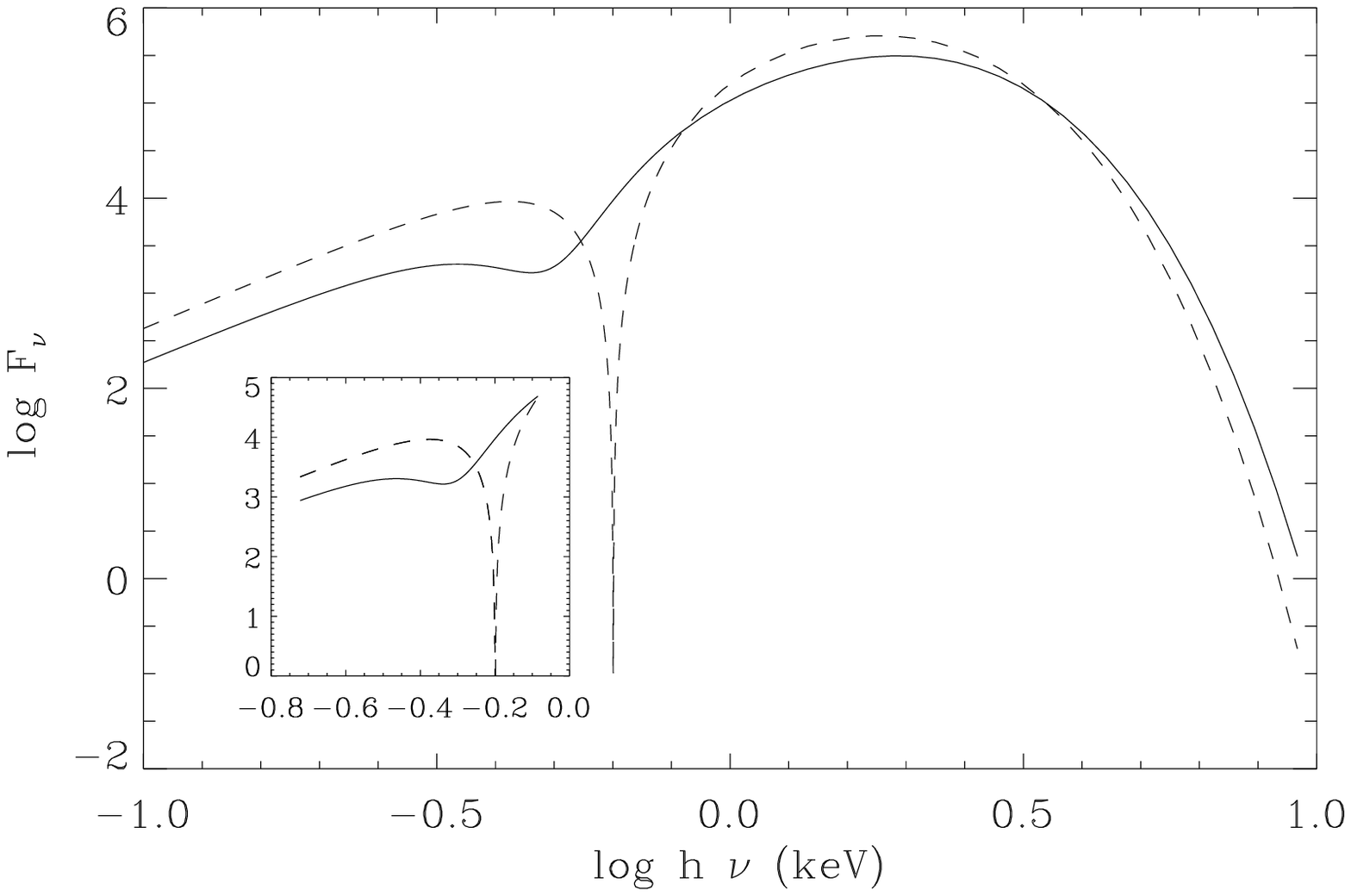}}}}

\caption{Broadening of the proton cyclotron line for
$B=10^{14}$~G and different luminosities. From left to right:
$L =  10^{33},  10^{34},  10^{35} \ {\rm erg\,
s}^{-1}$;
solid lines correspond to a
dipolar field, and dash-dotted lines to
$B$ normal to the surface. Flux is in arbitrary units and energies
have not been red-shifted at Earth.
\label{broad_14}}
\end{figure}

\begin{figure}

\centerline{
\epsfxsize=6.truecm
\centerline{{\epsfbox{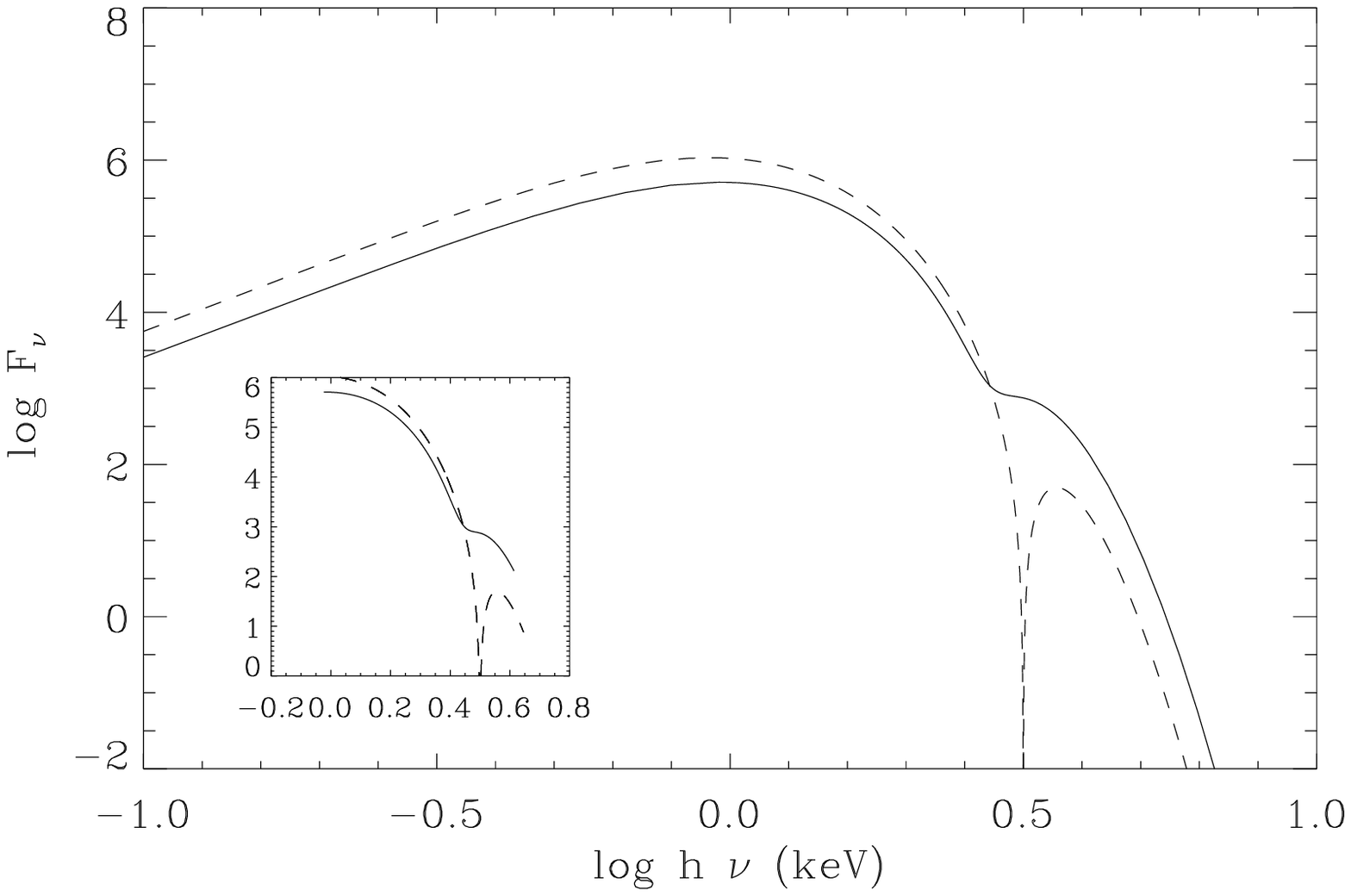}}
{\epsfbox{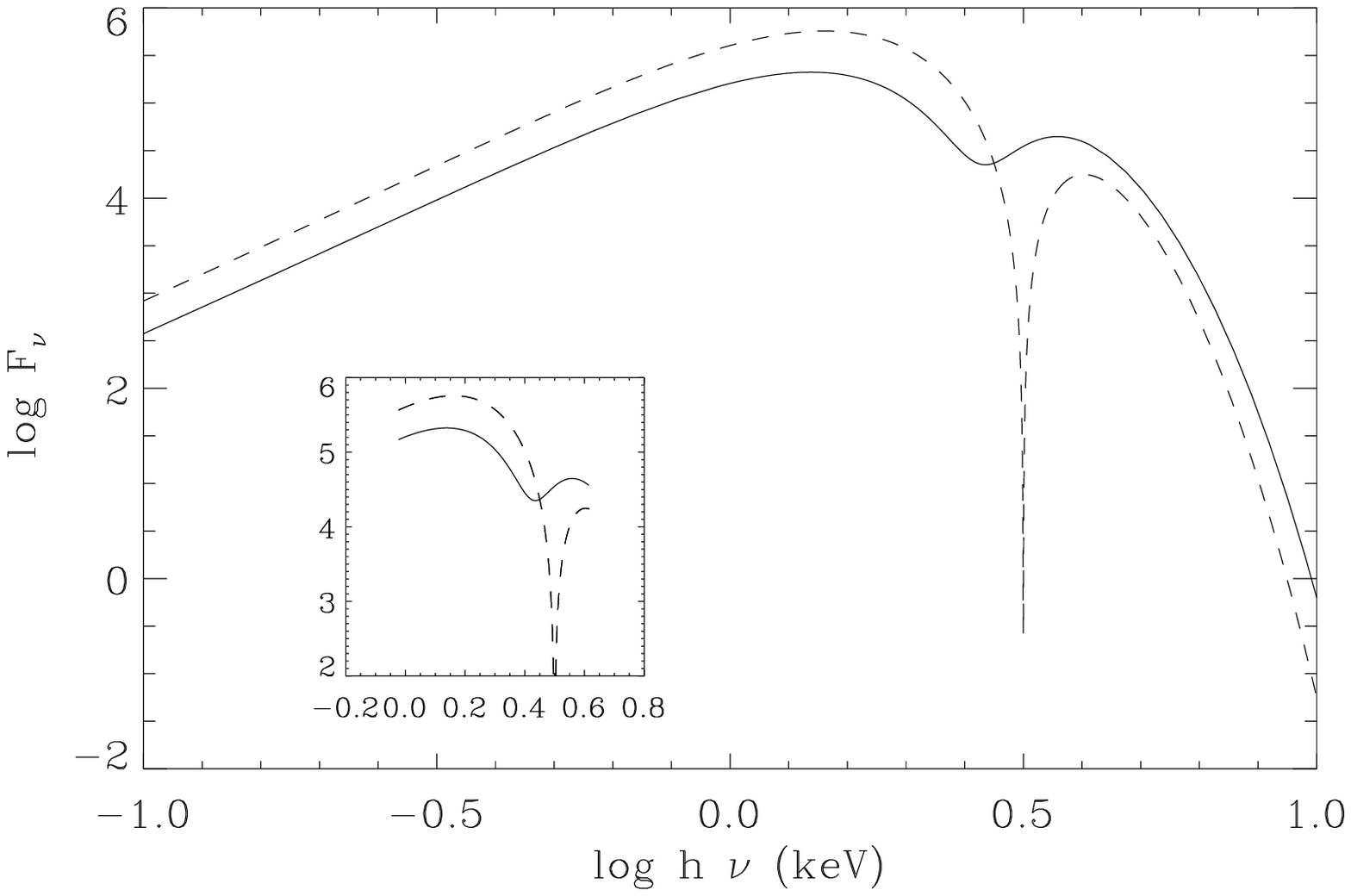}}
{\epsfbox{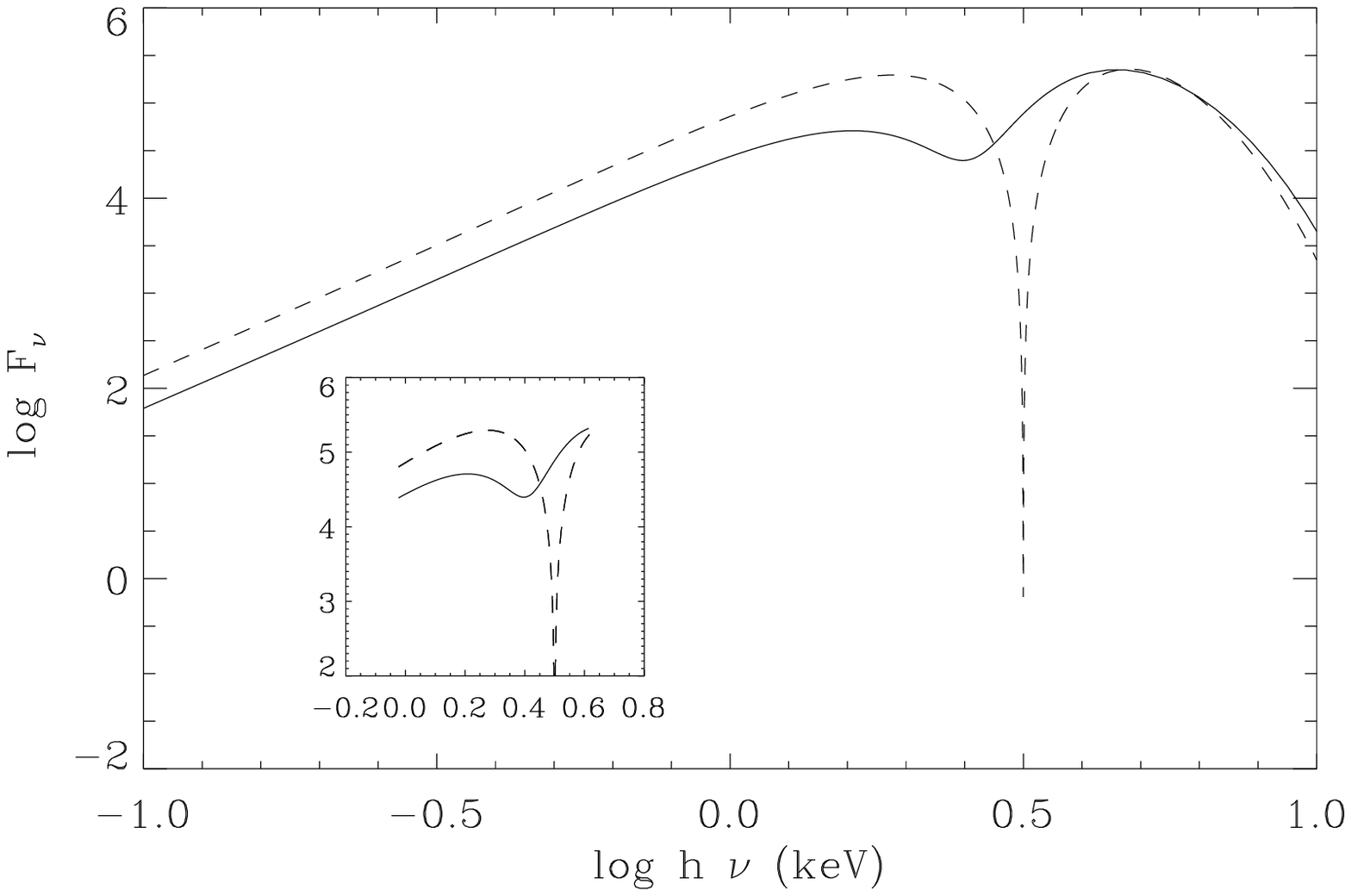}}}}
\caption{Same as in figure \ref{broad_14} for
$B=5 \times 10^{14}$~G and $L =  10^{34},  10^{35},  10^{36} \ {\rm erg\,
s}^{-1}$.
\label{broad_514}}
\end{figure}

\begin{figure}

\centerline{
\epsfxsize=6.truecm
\centerline{{\epsfbox{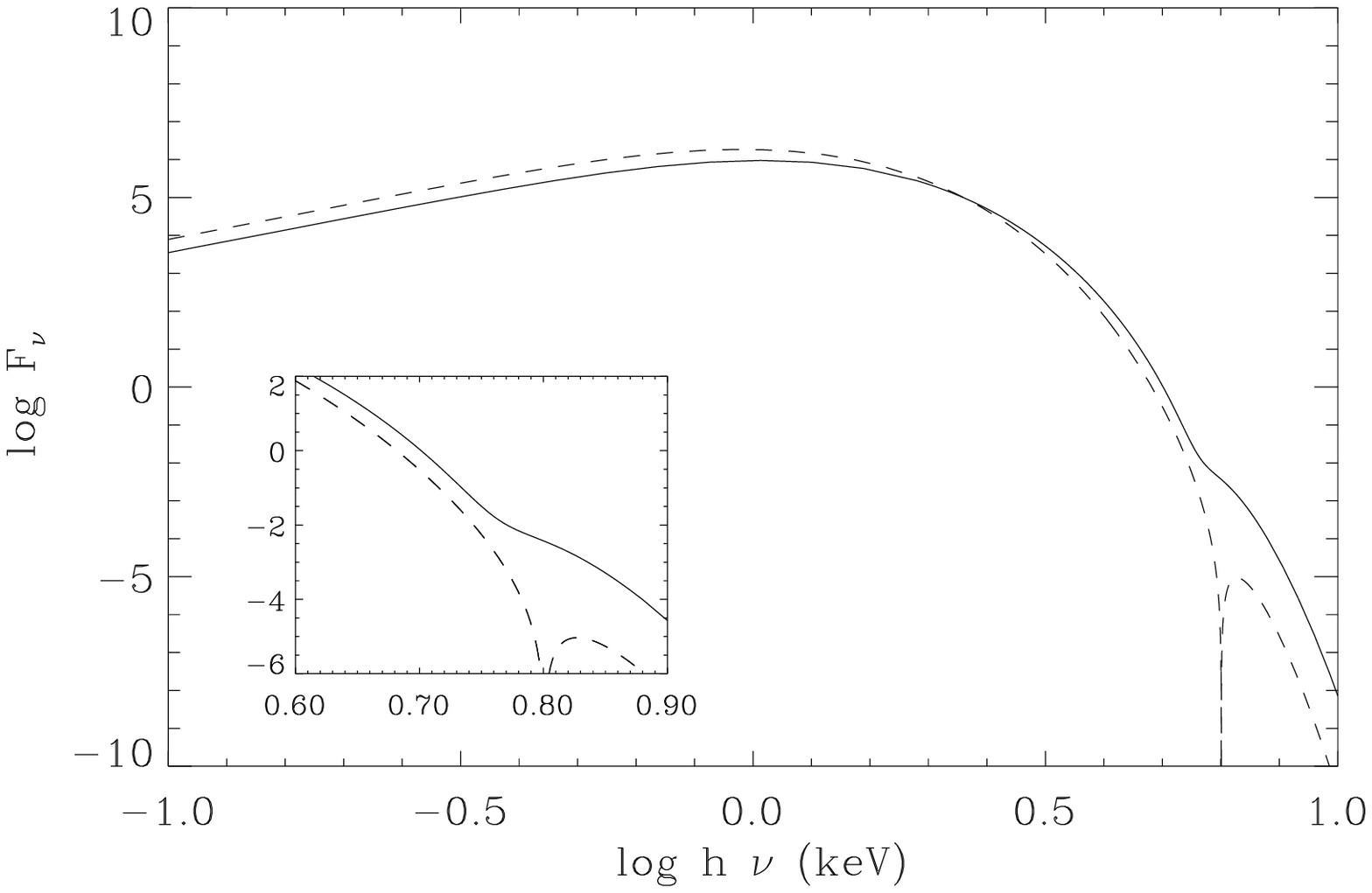}}
{\epsfbox{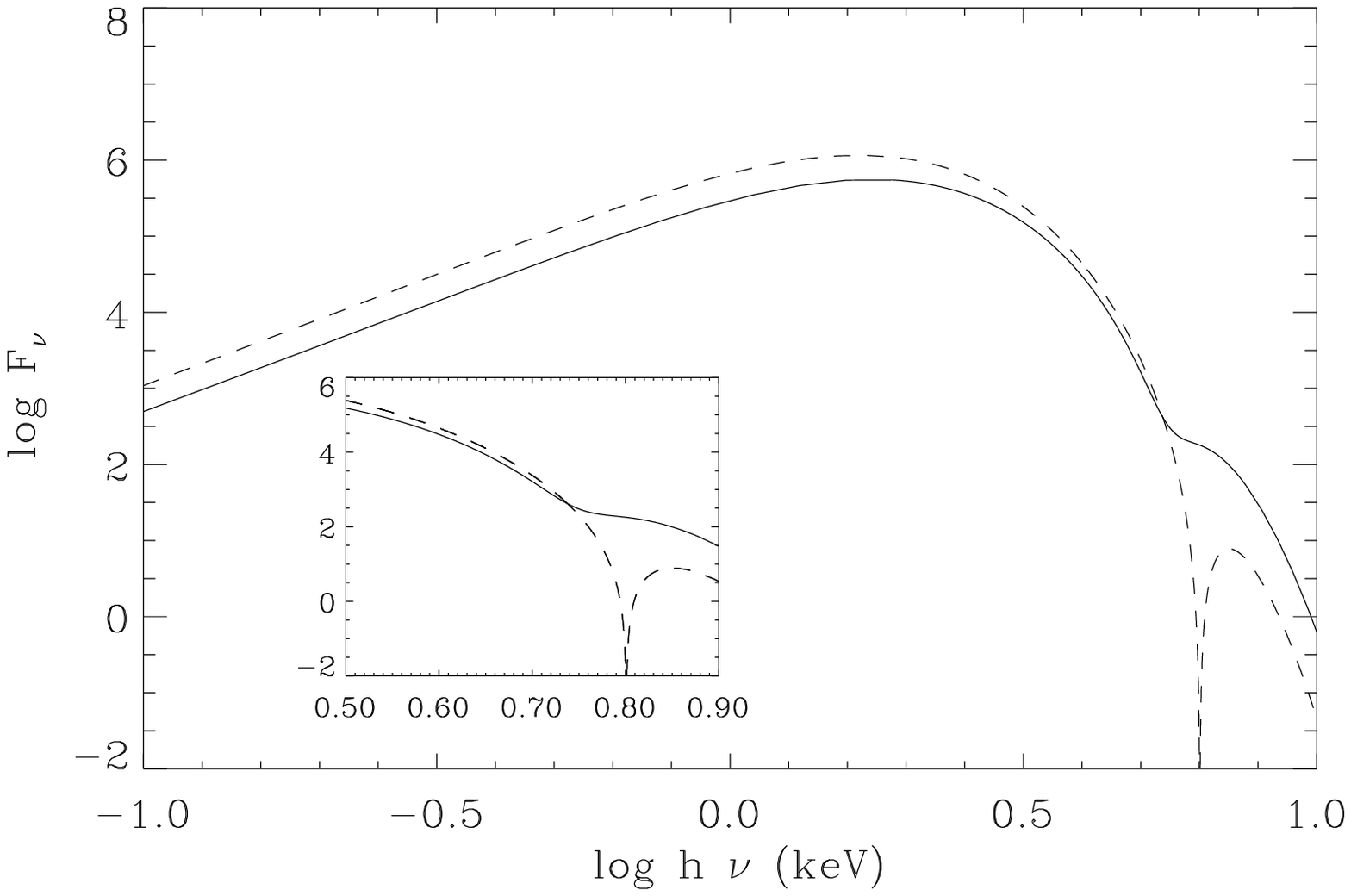}}
{\epsfbox{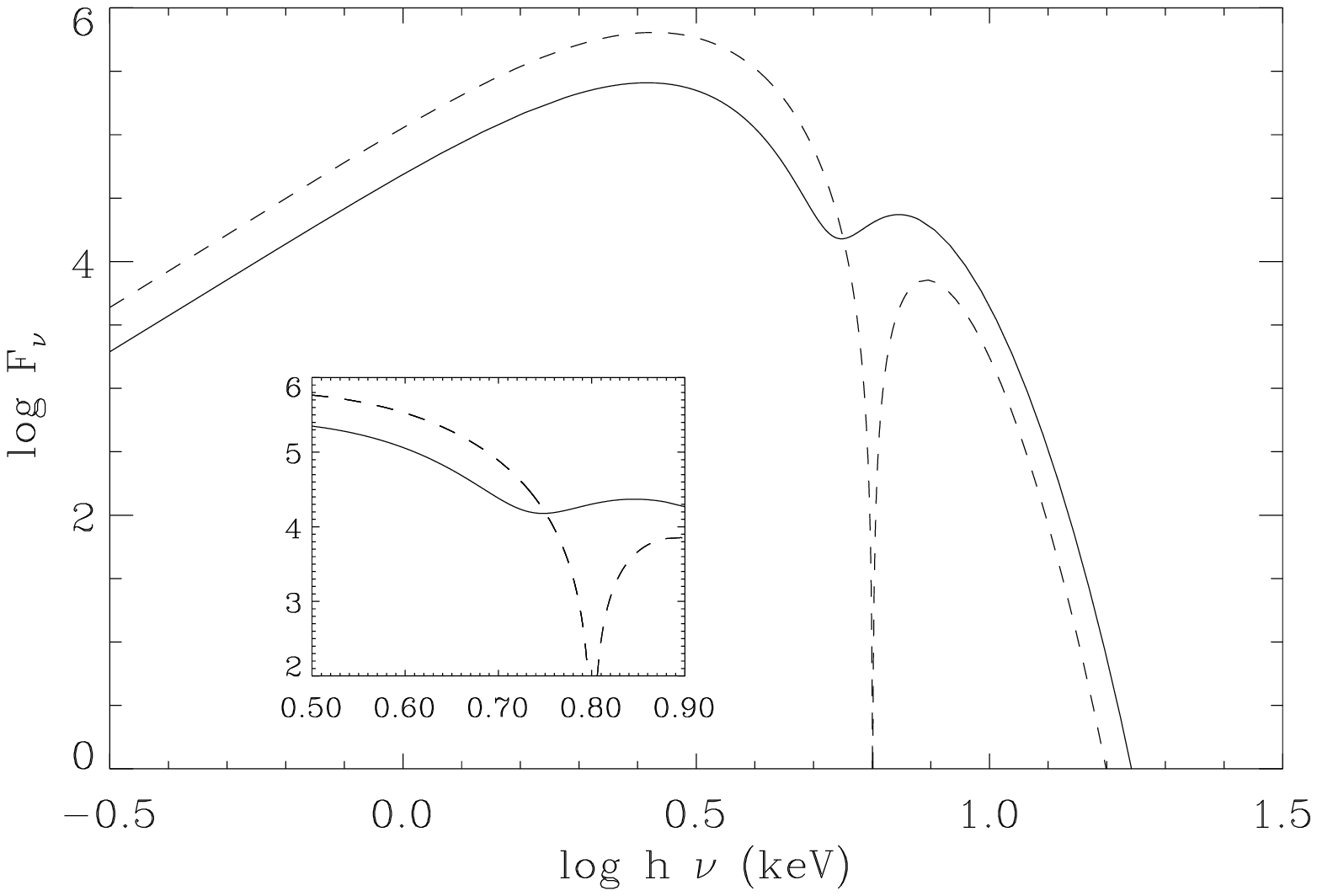}}}}
\caption{Same as in figure \ref{broad_514} for
$B= 10^{15}$~G.
\label{broad_15}}
\end{figure}

Line parameters derived from the previous calculation are
reported in table \ref{tab2} for different values of the luminosity
and of the polar $B$-field; diffusion-limit
spectra, as seen at the top of the NS atmosphere, are shown in figures
\ref{broad_14}-\ref{broad_15}. For $B_p=10^{15}$ G the PCF falls in
the Wien tail
and it is spread over a few keVs. To allow a direct comparison with the
case of a magnetic field parallel to the surface normal, we have repeated
the same calculation for a plane-parallel slab with constant $B=B_p$,
neglecting surface temperature variations. We stress that in both cases no mode
crossing effects have been accounted for, and, to be consistent
with the diffusion approximation, spectra were computed at optical
depth $\gtrsim 1$. Clearly these approximated results may be taken
only as indicative of the true shape of the PCF in emergent
spectra; it is however worth noticing that the line equivalent widths
derived from our simple analysis for ${\bf B}\, ||\, {\bf n}$ are in rough
agreement with those of the complete transfer calculation for the same
values of $B$ and $L$. As expected, the PCF turns out
to be systematically broader when emission comes from the entire star
surface, typically by 10--20\%. Also, the change of the field strength
with the latitude produces a shift of the line centroid toward lower
energies by $\sim$ 20--30\%. Both these effects are quite independent of
the values of $B_p$ and $L$ (see table \ref{tab2}). Complex magnetic
field
configurations with contributions from higher multipoles, such as those
envisaged in magnetar models (Thompson \cite{th00:2000}) or in the
modelling of isolated neutron stars (e.g. Geminga, Page et al.
\cite{pa95:1995}), perhaps represent a situation more realistic than the
dipolar field we have considered. Due to the steep
effective temperature gradients expected in this case, the real properties
of PCFs should be intermediate between those discussed above.

\begin{deluxetable}{ccccccc}
\tablecolumns{7}
\tablewidth{0pt}
\tablecaption{Line Broadening\label{tab2}}
\tablehead{
\colhead{$B_p$}&
\colhead{$L$} &
\colhead{$EW_{D}$} &
\colhead{$EW_{D}/EW_{||}$} &
\colhead{$E_{c,D}$} &
\colhead{$E_{c,D}/E_{c,||}$} &
\colhead{$\Delta E_{D}/E_{c,D}$} \\
\colhead{$10^{13}$ G} &
\colhead{$10^{34}$ erg s$^{-1}$} &
\colhead{keV} &
\colhead{} &
\colhead{keV} &
\colhead{} &
\colhead{}
}
\startdata
  1  & 0.1 & 0.035 & 1.17 & 0.046 & 0.73 & 0.84 \\
  1  & 1   & 0.035 & 1.20 & 0.046 & 0.73 & 0.84 \\
  1  & 10  & 0.035 & 1.22 & 0.047 & 0.75 & 0.80 \\
  10 & 0.1 & 0.35  & 1.14 & 0.49  & 0.78 & 0.75 \\
  10 & 1   & 0.35  & 1.12 & 0.47  & 0.75 & 0.86 \\
  10 & 10  & 0.34  & 1.08 & 0.46  & 0.73 & 0.73 \\
  50 & 1  & 1.74  & 1.06 & 2.57  & 0.81 & 0.53 \\
  50 & 10  & 1.72  & 1.06 & 2.72  & 0.86 & 0.63 \\
  50 & 100  & 1.75  & 1.11 & 2.49  & 0.79 & 0.79 \\
\tablecomments{
$D$ = dipolar field; $||$ = constant $B$ field. $E_c$ is the line
centroid and $\Delta E_{D}$ is the FWZI in the case of dipolar field;
energies have not been corrected for the gravitational red-shift.}
\enddata
\end{deluxetable}

\section{Discussion}\label{sec-discuss}

In this paper we have modeled thermal emission from
ultra-magnetized NSs, showing that the emergent continuum
is almost indistinguishable from a
blackbody at $T_{eff}$. We point out the existence of
a strong absorption feature at the proton cyclotron energy; the line
equivalent width is between $30$ eV and 1 keV and the line center is
located at
$\sim 0.5$--5 keV for field strengths $\sim 10^{14}$--$10^{15}$ G, which
are currently believed to be typical of SGRs and AXPs.
The radiative transfer calculations presented here corroborate the original
suggestion by Thompson (\cite{th00:2000})
that an emission/absorption feature
at the proton cyclotron resonance should be present in the thermal
spectrum of magnetars.

The natural width and the line-splitting of electron cyclotron lines
were thoroughly investigated by Pavlov et al. (\cite{pbma91:1991}) by means
of a rigorous QED treatment. No similar calculation is available for
protons, but, in any realistic situation,
$B\ll B_{crit,p}=m_p^2c^3/\hbar e \sim 1.5\times 10^{20}$ G,
so protons are never relativistic. In this case, the natural width of the
line associated with the transition between the first and the ground
Landau levels is $\Gamma^\pm_1\sim (4/3)\alpha m_pc^2 (B/B_{crit,p})^2$
(Pavlov, private communication), where $\alpha$ is the fine structure
constant and the plus/minus sign
refers to spin-up $\leftrightarrow$ spin-up, spin-down $\leftrightarrow$
spin-down transitions respectively.
In addition to the first order cyclotron resonance Thompson
(\cite{th00:2000}) also suggested the possibility to detect the spin-flip
line, despite the fact that no detailed calculation has been presented.
In fact, contrary to electrons, which have
a small anomalous magnetic moment $\Delta\mu/\mu_0\approx\alpha$ (here
$\mu_0$ is the B\"ohr magneton), $\Delta\mu/\mu_N\simeq 2.79$ for
protons, $\mu_N=(m_p/m_e)\mu_0$ being the nuclear magneton. This implies
that the proton spin-flip transition produces a line at $2.79E_{c,p}$ which
is always well-separated from the fundamental resonance
at $E_{c,p}$. Here we only note that this second line should be much
weaker and is probably beyond the capabilities of the present
spectrometers, because the rates for the spin-flip transitions in the
non-relativistic regime should contain an additional factor $\approx
E_{c,p}/m_pc^2 \ll 1$ with respect to transitions with conserved spin
projection (Pavlov, private
comunication; see also Melrose \& Zheleznyakov \cite{mz81:1981} for the
case of electrons).  On the other hand, the fact that the lowest Landau
level is always non-degenerate for protons might prove
decisive in discriminating between proton and electron
absorption features in high-resolution spectral observations.
In fact, if absorption features observed at $0.06-6$~keV
are interpreted as electron cyclotron resonances, the inferred magnetic
fields should be as low as $10^9-10^{11}$~G. For values in this
range, the interaction with the radiation field should remove the degeneracy
of the lowest Landau level, giving rise to a doublet.
In the case of electrons, even if the two lines are not individually
resolved (the required resolving power should be $\sim E_{c,e}/\Delta E
\sim (0.16 \alpha)^{-1} \sim 860$)
observations with a narrow
band Bragg polarimeter should show a sharp jump when scanning the feature
from the blue toward the red wing (see
Pavlov et al. \cite{pbma91:1991} for details).

Finally, we would like to stress that
present results have been derived under a set of simplifying
assumptions, on both the atmospheric properties and the line profiles.
Atmospheres around young NSs are likely to be rich in metals.
Unfortunately the absorption coefficients for metals in a strong magnetic
field are poorly known and our assumption of a pure H composition,
which is certainly inadequate, appears at present the only viable
option.
Metal lines may well be present in magnetars X-ray spectrum. If originating
in an accreting gas or a corona they are expected to be mainly in
emission and narrower, unless the circumstellar material where the line
originates has a velocity large enough to produce a (Doppler-) \newline
broadening
comparable to that of the PCF. This is not obvious if the NS atmosphere
itself is rich in heavy elements since energies and shapes of atomic
transitions are strongly dependent on the local conditions (see Rajagopal, \&
Romani \cite{rr96:1996}; Zavlin et al. \cite{zps96:1996}). Strongly
magnetized Fe spectra presented by Rajagopal, Romani, \& Miller (\cite{rrm97:1997})
show significant absorption features in the X-rays. A confusion between
metal lines and PCFs
is therefore possible and this issue should be carefully addressed when
observed spectra are analyzed. Moreover, for a different chemical composition, lower
energy cyclotron lines at the various ion resonances should also be
present, since $E_{c,i}= (Z/A)E_{c,p}$. A firm identification of an
observed feature in
terms of a PCF may follow from the detection of higher harmonics (already
observed in some pulsars for the electron cyclotron resonance), despite
they are expected to be weaker (but see Cosumano et al. \cite{co98:1998}
for a counterexample). In the normal mode approximation, higher-order harmonics
of the electron cyclotron
resonance can be accounted for in the Gaunt factors (see Pavlov \& Panov
\cite{pp76:1976}; these expressions have been used here and in ZTT), but
we are unaware of a similar treatment of the effective collision frequencies for
the higher harmonics of the proton resonance.

Partial H ionization should also be taken
into account.  Although no detailed investigation on the ionization
properties of hydrogen in ultra-strong fields have been carried out,
the results of Potekhin, Chabrier \& Shibanov (\cite{pcs99:1999}) for
$B\lesssim 10^{13}$ G
clearly indicate that the neutral fraction increases with the field
strength. For values of density and temperature typical of the inner layers
of our atmospheric models,
the neutral fraction is $\approx 0.01$ at $B=10^{13}$ G
and it is bounded to increase at larger $B$. This will reduce the number of
free protons, thus making resonant absorption less effective.

Line-broadening mechanisms have not been accounted for
in our present investigation, apart from the smearing of the line
introduced by
the dipolar variation of the $B$-field (see \S\ref{sec-results}).
The more important broadening could be due
to the thermal Doppler effect, which produces a width
$\Delta E_{th}\approx E_{c,p}(kT/m_p c^2)^{1/2} \approx 10^{-4}
E_{c,p}$ under typical conditions. As expected,
Doppler-broadening is less important by far for protons with respect to
electrons and $\Delta E_{th}\ll \Delta E$ in our models (see table \ref{tab2}).

Up to now no positive detection of any spectral feature, both in emission
and in absorption, have been reported in SGRs and AXPs
(see Mereghetti \cite{me00:2000} and references therein).
The only possible exception is the anomalous X-ray pulsar 1E 2259+586, for
which the detection of a cyclotron feature around 5 keV has been
claimed (Iwasawa, Koyama \& Halpern \cite{ikh92:1992}) but never confirmed
by subsequent observations (e.g. Corbett et al. \cite{co95:1995}). The
discrepancy between the spin-down age and the age of the SNR in this
source hints toward a higher value of the field in the past, but
the large line energy would imply a present field $\approx 10^{15}$
G, which seems difficult to reconcile with the pulsar evolutionary
track (Colpi,
Geppert \& Page \cite{cpd00:2000}). However, present data do not exclude the
possibility that in  SGRs and AXPs the magnetic field is a few $10^{14}$ G.
If this is the case, a cyclotron feature around
$\sim 1$ keV may well have escaped detection because these sources have
never been observed with sufficient spectral resolution at low energies.

\acknowledgments

We are grateful to G.G. Pavlov for several enlightening
comments on the radiative width of the proton
cyclotron line and the relative importance of the proton
spin-flip transition. Work partially supported by Italian MURST
under grant COFIN-98-02154100.

\newpage

\end{document}